\documentstyle[preprint,floats,epsfig,prl,aps]{revtex}

\tightenlines

\begin{document}

\draft
\title{How an Anomalous Cusp Bifurcates}
\author{Robert S. Maier${}^{1,2}$ and D.~L. Stein${}^{2,1}$}
\address{${}^{1}$Mathematics and ${}^{2}$Physics Departments, University of
Arizona, Tucson, Arizona 85721}

\maketitle

\begin{abstract}
We study the pattern of activated trajectories in a double well system
without detailed balance, in the weak noise limit.  The pattern may contain
cusps and other singular features, which are similar to the caustics of
geometrical optics.  Their presence is reflected in the quasipotential of
the system, much as phase transitions are reflected in the free energy of a
thermodynamic system.  By tuning system parameters, a cusp may be made to
coincide with the saddle point.  Such an anomalous cusp is analogous to a
nonclassical critical point.  We~derive a scaling law, and nonpolynomial
`equations of state', that govern its bifurcation into conventional cusps.
\end{abstract}

\pacs{PACS numbers: 05.40.-a, 05.70.Fh, 42.15.Dp}

\narrowtext

The {\em optimal trajectory\/} concept has been widely used in the theory
of noise-activated
transitions~\cite{MaierC,MaierF,MaierD,Dykman92,Dykman94,Maier97b,Maier99}.
In~the weak noise limit, when transitions between stable states become
exponentially rare, one or at most a few trajectories in the system state
space are singled~out as escape paths of least resistance.  Also, between
the energetically lowest stable state and any other state there are at~most
a few dominant activated trajectories.  Such optimal trajectories, which
are determined by a `least energy expended' or `least action' variational
principle, are experimentally observable~\cite{Maier97b,Maier99}.  In
systems that have the property of detailed balance, they are time-reversed
relaxational trajectories.  But in nonequilibrium systems, which lack
detailed balance, the optimal trajectory pattern extending from a stable
state may be more complicated.

\begin{figure}[b]
\centerline{\epsfig{file=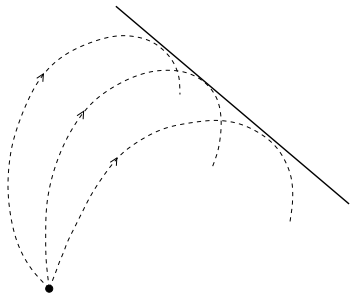,width=1.8in}}
\vspace{0.25in}
\caption{How a caustic (solid curve) could be formed as the envelope of
a family of outgoing optimal trajectories.}
\label{fig:caustic}
\end{figure}

Optimal trajectories are similar in many ways to the rays of geometrical
optics, which characterize in the short wavelength limit the waves
emanating from a point source.  That is because optical rays may be
computed variationally too, from a `least optical depth' principle.  In~a
medium with inhomogeneous index of refraction, it is common for a ray
family to bounce~off a {\em caustic surface\/}, leaving the region behind
in shadow: not illuminated, or illuminated only indirectly.  Other singular
features with a catastrophe-theoretic interpretation may be
produced~\cite{Berry76,Poston78}.  In~a noise-driven system in which
detailed balance is violated, the pattern of optimal trajectories may
contain similar features~\cite{MaierC,Dykman94,Jauslin87}.  See
Fig.~\ref{fig:caustic}.

To understand the crossing of optimal trajectories, it is useful to look at
the {\em quasipotential\/} of the noise-driven system.  If~$\epsilon$ is
the noise strength (e.g., $\epsilon=kT$ in thermal systems), and $\rho({\bf
x})$ denotes the stationary probability density of the system at
state~${\bf x}$, then a quasipotential $W=W({\bf x})$ may be defined
phenomenologically by
\begin{equation}
\label{eq:first}
\rho({\bf x}) \sim \text{const}\times\exp[-W({\bf x})/\epsilon],
\qquad \epsilon\to0.
\end{equation}
This definition makes sense whether or not the system dynamics are
conservative, and whether or not the noise acts so as to preserve detailed
balance.  $W$~equals zero at the energetically lowest stable state(s), and
$W({\bf x})$ is essentially the minimum energy needed to excite the system
to state~${\bf x}$.  It~may be computed as a line integral along the
optimal trajectory extending to~${\bf x}$.

Formally, $W$ is multivalued at any state, such as the states near a
caustic, that is reached by more than one optimal trajectory emanating from
the energetically lowest state(s).  But by~(\ref{eq:first}), the least
value is dominant, and the trajectories giving rise to others are
unphysical.  The state space of a noise-driven system is typically
partitioned by `switching surfaces', on which dominance switches between
branches of~$W$.

The switching of dominance resembles a first-order phase transition in a
condensed matter system.  The similarity is unsurprising, since phase
transitions with classical critical exponents also have a
catastrophe-theoretic interpretation~\cite{Poston78}.  Consider, for
example, a~ferromagnetic system with extensive order parameter~$m$
(magnetization), in a magnetic field~$h$.  Its thermodynamics are
determined by a free energy function $\Psi=\Psi(T,h)$.  Below the critical
temperature~$T_c$, $\Psi$~and $m=\partial\Psi/\partial h$ are multivalued.
If~the phase transition is classical, i.e., of mean-field form, $\Psi$~is
three-valued in a sharp-tipped region of the $(T,h)$ plane bounded by
`spinodals' of the form $h \approx \pm\text{const}\times (T_c -T)^{3/2}$.
That is because the leading terms in the Legendre transform
$\widetilde\Psi^{(h)}(T,m) \equiv hm-\Psi(T,h)$ are of Ginzburg--Landau
type:
\begin{equation}
\label{eq:GLferro}
\widetilde\Psi^{(h)}(T,m)\approx C_2 (T-T_c) {m^2}/2 + C_4 {m^4}/4.
\end{equation}
In~the catastrophe-theoretic sense~\cite{Berry76,Poston78}, the spinodals
are {\em fold caustics\/}.  Each is the projection of a fold in the graph
of~$\Psi$, which is a two-dimensional surface, onto the $(T,h)$ plane.  The
critical point $(T_c,0)$ from which the spinodals extend is a~{\em cusp
catastrophe\/}: the projection of the point on the graph of~$\Psi$ at~which
the two folds join.  A~first-order phase transition line, on which
dominance switches between branches of~$\Psi$, extends from $(0,0)$
to~$(T_c,0)$.

Switching lines in two-dimensional noise-driven systems are clearly
analogous to first-order phase transition lines, and caustics to spinodals.
Caustics typically terminate at cusps, and switching lines also frequently
terminate at cusps.  So cusps, which are very common, are analogous to
second-order critical points~\cite{MaierF}.  They are physically important
because at any cusp, the prefactor `$\text {const}$' in~(\ref{eq:first}),
which in~general is ${\bf x}$-dependent, diverges.

In previous work~\cite{MaierC,MaierF}, we pointed~out that in many
noise-driven two-dimensional double well systems without detailed balance,
a~cusp may be moved to coincide with the saddle point between the wells, by
tuning parameters.  If they coincide, the Kramers (${\epsilon\to0}$) limit
of noise-induced interwell transitions is greatly affected.  The prefactor
in the Kramers transition rate formula becomes anomalous: it~acquires a
negative power of~$\epsilon$.

Precisely at criticality, we were able to approximate the
quasipotential~$W$ near any such `anomalous cusp'.  Our expression differed
from the polynomial `normal forms' of conventional catastrophe theory.
In~thermodynamics, it would correspond to a {\em nonclassical\/} phase
transition.  In~catastrophe theory, it would be interpreted as a {\em
nongeneric\/} catastrophe: one of the few such of physical relevance to
have been discovered since the work of Berry and Mount on the short
wavelength limit of scattering~\cite{Berry76}.

In~this Letter, we extend Ref.~\cite{MaierF} by analysing the `unfolding'
of an anomalous cusp in a typical two-dimensional noise-activated system,
as a parameter is moved toward or away from criticality.  We explain how it
may bifurcate into conventional cusps.  Our scaling law for the bifurcation
yields a corresponding law for the divergence of the Kramers
prefactor~\cite{MaierSteininpreparation}.

Consider the following double well model, which is similar to models of
blocking dynamics in glassy systems, where a particle is coupled to a
randomly fluctuating barrier whose position is coupled to the particle
motion~\cite{Stein89}.  Let $x$ (a~particle position variable) and~$y$
(a~barrier state variable) be nontrivially coupled in such a way that the
values $\pm1$ for~$x$ and $0$~for~$y$ are stable.  If $x$ and~$y$ are
overdamped and are driven by white noise of strength~$\epsilon$, their
joint dynamics could be modeled by Langevin equations
\begin{eqnarray}
\dot x & = & \lambda_x \left [x \left(1 - x^2\right) - \alpha xy^2
\right ]+ \epsilon^{1/2}\eta_x(t) \nonumber\\ 
\dot y & = & -\mathopen|\lambda_y\mathclose| 
(1 + x^2) y + \epsilon^{1/2}\eta_y(t).
\label{eq:driftfield}
\end{eqnarray}
The parameters $\lambda_x>0$ and $\lambda_y<0$ determine the time scales on
which $x$ and~$y$ evolve, and govern the all-important relaxational
behavior near the saddle point $(0,0)$, where $(\dot x,\dot y) \approx
(\lambda_x x, -\mathopen|\lambda_y\mathclose|y)$.  The forcing terms
$(\eta_x,\eta_y)$ are a pair of independent Gaussian white noises, so that
$\langle \eta_i(s)\eta_j(t) \rangle$ equals $\delta_{ij}\delta(s-t)$.  The
parameter~$\alpha$ controls the absence of detailed balance: only when
$\alpha$ equals $\mu\equiv\mathopen|\lambda_y\mathclose|/\lambda_x$ is
there detailed balance, since only in that case is the drift field derived
from a potential.

Our results are insensitive to the details of the coupling between $x$
and~$y$, so long as the model is symmetric through $x=0$ and $y=0$.
To~compute the pattern of optimal trajectories emanating from the bottom of
the $x<0$ well or the $x>0$ well, we~use the fact that in any
multidimensional noise-driven system with vector Langevin equation $\dot
{\bf x} = {\bf u}({\bf x}) + {\epsilon}^{1/2}{\bbox{\eta}}(t)$, the optimal
trajectories are really zero-energy {\em Hamiltonian\/} trajectories,
generated by the Wentzell--Freidlin Hamiltonian~\cite{VF}
\begin{equation}
\label{eq:Hamiltonian}
H({\bf x},{\bf p}) ={\bf p}^2/2 + {\bf u}({\bf x})\cdot{\bf p}.
\end{equation}
That is because the associated Hamilton's principle is
\begin{displaymath}
\delta \int L({\bf x},\dot{\bf x})\,dt =
\delta \int \left|\dot {\bf x} - {\bf u}({\bf x})\right|^2\,dt
= 0,
\end{displaymath}
which is clearly a `least energy expended' principle.  The conjugate
momentum ${\bf p}\equiv \partial L/\partial \dot {\bf x}$ equals $\dot {\bf
x} - {\bf u}$, which measures the system's motion against the drift.

\begin{figure}[t]
\centerline{\epsfig{file=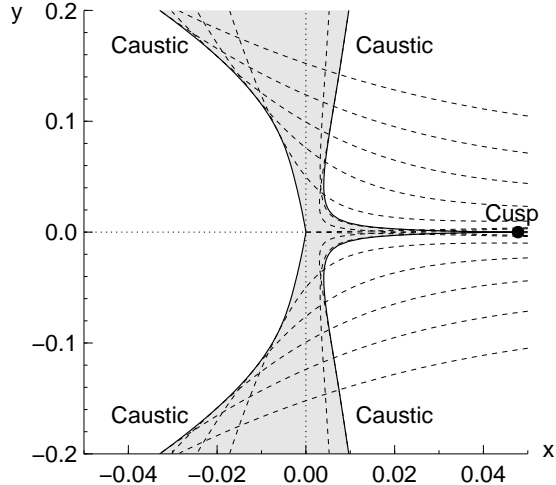,width=2.85in}}
\vspace{0.25in}
\caption{The pattern of optimal trajectories near the saddle, in the
`broken phase' ($\alpha>\alpha_c$).  Here $\lambda_x=1$,
$\mathopen|\lambda_y\mathclose|=1.1$, $\alpha=4.64$, and the critical value
$\alpha_c$ equals $4.62$.  In~the shaded region the quasipotential $W$ is
multivalued, and a switching line extends from $(0,0)$ to the cusp
$(x_c,0)\approx (0.048,0)$.}
\label{fig:broken}
\end{figure}

Figure~\ref{fig:broken} was obtained from (\ref{eq:driftfield})
and~(\ref{eq:Hamiltonian}) by integrating Hamilton's equations outward,
at~zero energy, from $(1,0)$, i.e., from the bottom of the right-hand well.
A~small portion of the $x<0$ half-plane is reached by optimal trajectories,
but the rest is in shadow.  In~phase space, which is four-dimensional, the
optimal trajectories trace~out a two-dimensional manifold, called a
Lagrangian manifold.  This manifold lies `above' only a small part of the
$x<0$ half-plane.  It~folds over, covering the shaded portion of the
$(x,y)$ plane more than once.  In~the shaded region, the momentum~$\bf p$
and the quasipotential~$W$, which equals $\smallint {\bf p}\cdot d{\bf x}$,
are two-valued ($x<0$) or three-valued ($x>0$).  At any point~$\bf x$, $\bf
p$~equals $\bbox{\nabla}W({\bf x})$.

In Fig.~\ref{fig:broken}, the parameter $\alpha$ is chosen to be slightly
greater than a certain critical value,~$\alpha_c$.  If~there is detailed
balance, the optimal trajectory pattern contains no singular features, but
if $\alpha$~is increased through~$\alpha_c$, a~cusp $(x_c,0)$ emerges from
the saddle point at~$(0,0)$ and moves toward~$(1,0)$.  This phenomenon is
not peculiar to the model defined by~(\ref{eq:driftfield}).  In~any
symmetric two-dimensional double well system that violates detailed balance
and has a tunable parameter, a~similar cusp may be born.  The focusing of
optimal trajectories at the cusp resembles the focusing of rays in a
radially symmetric optical system.

The value~$\alpha_c$ can be computed from the second-order variational
equation $\delta^2 \smallint L\,dt=0$, which is a criterion for
bifurcation.  On physical grounds, when $\alpha<\alpha_c$, $\delta^2
\smallint L\,dt$~computed along the on-axis trajectory to the saddle is
positive, but when $\alpha>\alpha_c$, it~is negative.  In the
model~(\ref{eq:driftfield}), $\alpha_c$ turns~out (cf.\ Ref.~\cite{MaierF})
to equal $2\mu(\mu+1)$.

The cusp $(x_c,0)$ that is present when $\alpha>\alpha_c$ resembles a
second-order critical point.  $W$~is three-valued in the sharp-tipped
region extending from~it, which is bounded by `spinodals' of the form
$y\approx\pm\text{const}\times(x_c-x)^{3/2}$.  (See Fig.~\ref{fig:broken}.)
Moreover, there is a switching line extending from the saddle at~$(0,0)$ to
$(x_c,0)$.  As~noted, this line is analogous to a first-order phase
transition line.

What remains to be understood is how the cusp is born at~$\alpha=\alpha_c$.
In a three-dimensional space with coordinates $(x,y;\alpha)$, there is a
line of second-order critical points in the $y=0$ plane that extends from
$(0,0;\alpha_c)$ to~$(1,0;+\infty)$.  By~analogy with thermodynamics, one
might expect $(0,0;\alpha_c)$ to be a third-order critical point.  At~any
fixed $\alpha>\alpha_c$, the leading terms in the Legendre transform
$\widetilde W^{(y)}(x,p_y) \equiv yp_y - W$, close to the cusp, are
known to be of Ginzburg--Landau type~\cite{MaierF,Dykman94}:
\begin{equation}
\widetilde W^{(y)}(x,p_y)\approx
C_2(\alpha) \bigl[x-x_c(\alpha)\bigr] {p_y^2}/2 + C_4(\alpha) {p_y^4}/4.
\end{equation}
One might expect that the correct three-dimensional generalization would be
a higher-degree polynomial in $x$, $p_y$, and $\alpha-\alpha_c$.  That
would allow the birth of the cusp to be viewed as a classical phase
transition, or one of the generic (polynomial) elementary
catastrophes~\cite{Berry76,Poston78}.

In Ref.~\cite{MaierF} we presented initial evidence against this.
At~criticality ($\alpha=\alpha_c$), we were able to construct a scaling
solution for~$W$, valid near the $x$-axis close to the saddle.  The
equation satisfied by the scaling function contained non-integer powers:
in~fact, powers that depended continuously on the model parameter~$\mu$.

By linearizing Hamiltonian dynamics near the saddle, we have now
characterized fully the behavior of the quasipotential~$W$ near a singular
point like $(0,0;\alpha_c)$.  Our chief new result is a {\em cubic
equation\/} satisfied by the double Legendre transform of~$W$\null.
It~defines a higher-order, but nonclassical, critical point.  We~have also
extended our $\alpha=\alpha_c$ scaling law to the case when
$\mathopen|\alpha-\alpha_c\mathclose|$ is nonzero but small.  These results
should extend to any symmetric double well system with a tunable parameter.

\begin{figure}[t]
\centerline{\epsfig{file=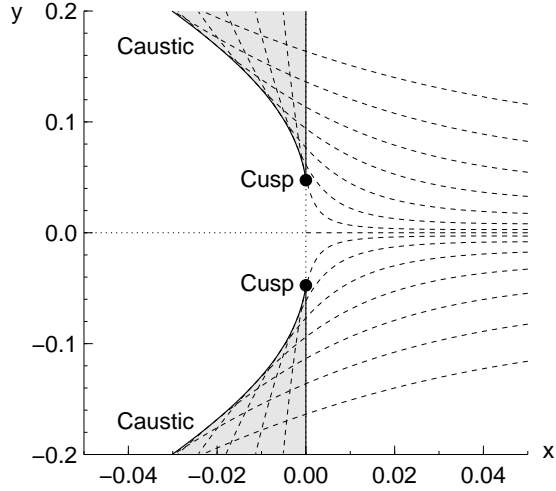,width=2.85in}}
\vspace{0.25in}
\caption{The pattern of optimal trajectories near the saddle, in the
`unbroken phase' ($\alpha<\alpha_c$).  Here $\lambda_x=1$,
$\mathopen|\lambda_y\mathclose|=1.1$, $\alpha=4.61$, and the critical value
$\alpha_c$ equals $4.62$.  As $\alpha\to{\alpha_c}^-$, the two cusps close
to the $y$-axis neck down to the saddle.}
\label{fig:unbroken}
\end{figure}

Figure~\ref{fig:unbroken}, which was obtained in the `unbroken phase'
($\alpha<\alpha_c$), sheds light on behavior near criticality.  The crucial
feature is the two caustics, relics of which appeared in
Fig.~\ref{fig:broken}.  They form part of the boundary of the `illuminated'
region.  Each caustic extends from a cusp, which is located very close to
the $y$-axis separatrix between the two wells.  As~$\alpha\to{\alpha_c}^-$,
the cusps {\em neck down\/} to the saddle.  Any further increase
in~$\alpha$ causes the $x$-axis cusp to be born, and to move toward
positive~$x$.

The merged cusp at~$(0,0)$, when $\alpha=\alpha_c$, is truly anomalous.
It~is the projection of a point on the boundary of the Lagrangian manifold,
rather than of a point in its interior.  So it is a {\em boundary
catastrophe\/}: a~singular point of a sensitive kind.  To explain its
bifurcation into cusps on the $x$-axis or $y$-axis, we must approximate the
optimal trajectory pattern in a neighborhood of the saddle, for
$\mathopen|\alpha-\alpha_c\mathclose|$ nonzero but small.

We accordingly linearize Hamilton's equations, which any optimal trajectory
must satisfy, near the point $(x,y;p_x,p_y)=(0,0;0,0)$ in phase space.
A~simple analysis (cf.\ Ref.~\cite{MaierD}) shows that this fixed point has
two stable directions, ${{\bf e}_s}=(0,1;0,0)$ and ${\tilde{\bf
e}_s}=(1,0;-2\lambda_x,0)$, and two unstable directions, ${{\bf
e}_u}=(1,0;0,0)$ and ${\tilde{\bf
e}_u}=(0,1;0,2\mathopen|\lambda_y\mathclose|)$.  The zero-momentum
directions (no~tilde) are eigendirections for relaxational trajectories,
which follow the drift.  In~the linear approximation, any optimal
trajectory near $(x,y)=(0,0)$ must satisfy
\begin{eqnarray}
\label{eq:linearized}
(x,y;p_x,p_y)&\approx&C_se^{-\mathopen|\lambda_y\mathclose|t}{{\bf
e}_s}+{{\widetilde C}_s} e^{-\lambda_xt}{\tilde{\bf e}_s} \\
&&{}+C_ue^{\lambda_xt}{{\bf e}_u} +{{\widetilde
C}_u}e^{\mathopen|\lambda_y\mathclose|t}{\tilde{\bf e}_u}\ , \nonumber
\end{eqnarray}
where the $C$'s are trajectory-specific constants.

We now index the `fan' of optimal trajectories that approach the saddle
point, as in Figs.\ \ref{fig:broken} and~\ref{fig:unbroken}, by~$s$.  The
normalization of this index variable is somewhat arbitrary.  A~reasonable
choice would be for it to denote distance from the $x$-axis (at~a fixed
$x>0$, near the saddle).  With this choice, $s=0$ will correspond to the
uphill optimal trajectory that climbs toward $(0,0)$ along the positive
$x$-axis.  If~each coefficient in~(\ref{eq:linearized}) can be expanded
in~$s$ about $s=0$, then by symmetry considerations
\begin{mathletters}
\label{eq:foureqns}
\begin{eqnarray}
C_s&=&a_1s+a_3s^3+\cdots,\\
{{\widetilde C}_s}&=&b_0+b_2s^2+\cdots,\\
C_u&=&c_2s^2+c_4s^4+\cdots,\\
{{\widetilde C}_u}&=&d_1s+d_3s^3+\cdots.
\end{eqnarray}
\end{mathletters}
We identify the passage through criticality, as $\alpha$ is increased
through~$\alpha_c$, with the passing through zero of the
coefficients $c_2$ and~$d_1$.  So, setting $\delta\equiv\alpha_c-\alpha$, we
take $c_2$ and~$d_1$ to be linearly proportional to~$\delta$, to leading
order.

Eq.~(\ref{eq:linearized}) comprises four scalar equations.  Eliminating $s$
and~$t$ among them, we can derive `equations of state' relating $\delta$
and any three of the phase space coordinates $x$, $y$, $p_x$, and~$p_y$.
When $\mathopen|\delta\mathclose|\ll1$, the equation relating $x$, $y$,
$p_y$, and~$\delta$ (near the $x$-axis), and the equation relating $x$,
$y$, $p_x$, and~$\delta$ (near the $y$-axis), turn~out to be, respectively,
\begin{mathletters}
\begin{eqnarray}
0&=&(p_y-2\mathopen|\lambda_y\mathclose|y)^3 \label{eq:eos1} \\
&&\quad {} +k_1\delta
x^{2\mu}(p_y-2\mathopen|\lambda_y\mathclose|y)+k_0x^{4\mu}p_y \nonumber \\
0&=&(p_x+2\lambda_x x)^3 \label{eq:eos2} \\
&&\quad {} +\ell_1\delta p_x^{2\mu-2}y^2(p_x+2\lambda_x x)
+\ell_0y^4p_x^{4\mu-3} \nonumber
\end{eqnarray}
\end{mathletters}
Here $k_1$, $k_0$, $\ell_1$, and~$\ell_0$ are positive constants.
At~criticality (${\delta=0}$), (\ref{eq:eos1})--(\ref{eq:eos2}) reduce to
the equations we previously obtained by an altogether different
technique~\cite{MaierF}.

By definition, the cusp $(x_c,0)$ is the point on the positive $x$-axis
where $W$ or~$(p_x,p_y)=\bbox{\nabla}W$ stops being multivalued as a
function of~$(x,y)$, as $x$~increases from~$0$.  It~is easy to verify
from~(\ref{eq:eos1}) that when $\delta$~is small and negative (i.e.,
$\alpha-\alpha_c$ is small and positive), the cusp location $x_c$ satisfies
$x_c\propto(-\delta)^{1/2\mu}$.  The $\delta$-dependence of the parent
cusps, which have $y=\pm y_c$, can be computed from~(\ref{eq:eos2}).  They
are the points close to the $y$-axis where $(p_x,p_y)$ first becomes
multivalued, as $\mathopen|y\mathclose|$ increases from zero.  When
$\delta\to0^+$ (i.e., $\alpha\to{\alpha_c}^-$), we~find that the parent
cusps neck down at the rate $y_c\propto\delta^{3/2-\mu}$.  That is so when
$1\le\mu<3/2$, at~least; for other~$\mu$, the prediction is that there are
no parent cusps, and no necking down.  All these predictions have been
numerically confirmed~\cite{MaierSteininpreparation}.

Despite the continuously varying exponents, the emergence of the $x$-axis
cusp is surprisingly similar to a second-order phase transition.  Recall
that close to the ferromagnetic critical point defined
by~(\ref{eq:GLferro}), scaled magnetization~$M$ and scaled magnetic
field~$H$ are related by
\begin{equation}
\label{eq:normalized}
M^3 \pm M - H = 0,
\end{equation}
or equivalently by the scaling law $M=\phi_\pm(H)$.  Here $M\equiv
Am/\mathopen|T-T_c\mathclose|^{1/2}$ and $H\equiv
Bh/\mathopen|T-T_c\mathclose|^{3/2}$, with $A\equiv\sqrt{C_4/C_2}$ and
$B\equiv A^3/C_4$.  The plus (minus) applies when $T-T_c$ is positive
(negative).  Eq.~(\ref{eq:eos1}) may be rewritten in the
form~(\ref{eq:normalized}), provided that one defines
\begin{eqnarray}
M&\equiv& (p_y - 2\mathopen|\lambda_y\mathclose|y) \big/ 
     \bigl|k_1\delta x^{2\mu}+k_0x^{4\mu}\bigr|^{1/2}\\
H&\equiv& 2\mathopen|\lambda_y\mathclose|k_0 x^{4\mu}y \big/
     \bigl|k_1\delta x^{2\mu}+k_0x^{4\mu}\bigr|^{3/2}.
\end{eqnarray}
The plus (minus) applies when $k_1\delta x^{2\mu}+k_0x^{4\mu}$ is positive
(negative).  The law $M=\phi_\pm(H)$, in which $\delta$ appears implicitly,
provides a unified description of the $x$-axis behavior both at
criticality ($\delta=0$) and away ($\delta\neq0$).

The most striking consequence of this approach is a general scaling law,
showing how the quasipotential varies as $(x,y;\alpha)$ moves away
from~$(0,0;\alpha_c)$, {\em in any direction\/}.  It~can be written using
the double Legendre transform $\widetilde W^{(x,y)}(p_x,p_y) \equiv {{\bf
x}\cdot{\bf p}-W}$, which equals $\smallint {\bf x}\cdot d{\bf p}$.  Near
the saddle,
\begin{eqnarray}
W(x,y)&\approx& W(0,0) - \lambda_x x^2 + \mathopen|\lambda_y\mathclose|y^2 \\
\widetilde W^{(x,y)}(p_x,p_y)&\approx& -W(0,0) -  p_x^2/4\lambda_x + p_y^2/4\mathopen|\lambda_y\mathclose|.
\label{eq:double}
\end{eqnarray}
Let $R=R(p_x,p_y)$ denote the difference between $\widetilde W^{(x,y)}$ and
the right-hand side of~(\ref{eq:double}).  Since $\widetilde W^{(x,y)}=
\smallint {\bf x}\cdot \dot{\bf p}\,dt$, $R$~may be expressed in terms of
$s$ and~$t$ by employing (\ref{eq:linearized}) and~(\ref{eq:foureqns}).  By
eliminating $s$ and~$t$ from the formul\ae\ for $R(s,t)$, $p_x(s,t)$, and
$p_y(s,t)$, we find to leading order in~$\delta$
\begin{equation}
\label{eq:cubic}
R^3 + m_1\delta p_x^{2\mu} p_y^2 R - m_0p_x^{4\mu}p_y^4 = 0,
\end{equation}
where $m_1$ and~$m_0$ are positive constants.  This is the extension to
$\delta\neq0$ of a formula derived in Ref.~\cite{MaierF}.

In phase transition language, the cubic equation~(\ref{eq:cubic}) fully
characterizes the nonclassical structure of the critical point
$(x,y;\alpha)=(0,0;\alpha_c)$, i.e., $(p_x,p_y;\alpha)=(0,0;\alpha_c)$.
Equation~(\ref{eq:cubic}) is also interesting from a catastrophe theory
point of view.  Nongeneric catastrophes, which are difficult to classify,
may in~general be perturbed in an infinite number of ways so as to yield
generic catastrophes~\cite{Berry76}.  But (\ref{eq:cubic}) describes the
bifurcation of a nongeneric catastrophe (the anomalous cusp) into
conventional cusps, in a unique, physically determined way.

This research was supported in part by NSF grant PHY-9800979.  A~portion
was completed while the authors were in residence at the Aspen Center for
Physics.


\end{document}